\documentstyle{mn}

\newif\ifAMStwofonts

\ifoldfss
  \ifCUPmtlplainloaded \else
    \NewTextAlphabet{textbfit} {cmbxti10} {}
    \NewTextAlphabet{textbfss} {cmssbx10} {}
    \NewMathAlphabet{mathbfit} {cmbxti10} {} 
    \NewMathAlphabet{mathbfss} {cmssbx10} {} 
  \fi
  \ifAMStwofonts
    \ifCUPmtlplainloaded \else
      \NewSymbolFont{upmath} {eurm10}
      \NewSymbolFont{AMSa} {msam10}
      \NewMathSymbol{\upi}     {0}{upmath}{19}
      \NewMathSymbol{\umu}     {0}{upmath}{16}
      \NewMathSymbol{\upartial}{0}{upmath}{40}
      \NewMathSymbol{\leqslant}{3}{AMSa}{36}
      \NewMathSymbol{\geqslant}{3}{AMSa}{3E}

      \let\leq=\leqslant 
       
    \fi
  \fi
\fi 

\ifnfssone
  \newmathalphabet{\mathit}
  \addtoversion{normal}{\mathit}{cmr}{m}{it}
  \addtoversion{bold}{\mathit}{cmr}{bx}{it}
  \newmathalphabet{\mathbfit} 
  \addtoversion{normal}{\mathbfit}{cmr}{bx}{it}
  \addtoversion{bold}{\mathbfit}{cmr}{bx}{it}
  \newmathalphabet{\mathbfss} 
  \addtoversion{normal}{\mathbfss}{cmss}{bx}{n}
  \addtoversion{bold}{\mathbfss}{cmss}{bx}{n}
  \ifAMStwofonts
    \ifCUPmtlplainloaded \else
      \UseAMStwoboldmath
      \makeatletter
      \new@mathgroup\upmath@group
      \define@mathgroup\mv@normal\upmath@group{eur}{m}{n}
      \define@mathgroup\mv@bold\upmath@group{eur}{b}{n}
      \edef\UPM{\hexnumber\upmath@group}
      \new@mathgroup\amsa@group
      \define@mathgroup\mv@normal\amsa@group{msa}{m}{n}
      \define@mathgroup\mv@bold\amsa@group{msa}{m}{n}
      \edef\AMSa{\hexnumber\amsa@group}
      \makeatother
      \mathchardef\upi="0\UPM19
      \mathchardef\umu="0\UPM16
      \mathchardef\upartial="0\UPM40
      \mathchardef\leqslant="3\AMSa36
      \mathchardef\geqslant="3\AMSa3E

      \let\leq=\leqslant 

    \fi
  \fi
\fi 

\ifnfsstwo
  \DeclareMathAlphabet{\mathbfit}{OT1}{cmr}{bx}{it}
  \SetMathAlphabet\mathbfit{bold}{OT1}{cmr}{bx}{it}
  \DeclareMathAlphabet{\mathbfss}{OT1}{cmss}{bx}{n}
  \SetMathAlphabet\mathbfss{bold}{OT1}{cmss}{bx}{n}
  \ifAMStwofonts
    \ifCUPmtlplainloaded \else
      \DeclareSymbolFont{UPM}{U}{eur}{m}{n}
      \SetSymbolFont{UPM}{bold}{U}{eur}{b}{n}
      \DeclareSymbolFont{AMSa}{U}{msa}{m}{n}
      \DeclareMathSymbol{\upi}{0}{UPM}{"19}
      \DeclareMathSymbol{\umu}{0}{UPM}{"16}
      \DeclareMathSymbol{\upartial}{0}{UPM}{"40}
      \DeclareMathSymbol{\leqslant}{3}{AMSa}{"36}
      \DeclareMathSymbol{\geqslant}{3}{AMSa}{"3E}

      \let\leq=\leqslant 

    \fi
  \fi
\fi 

\ifCUPmtlplainloaded \else
  \ifAMStwofonts \else 
    \def\upi{\pi}
    \def\umu{\mu}
    \def\upartial{\partial}
  \fi
\fi

\title[The effects of Moffat PSFs on S\'ersic profiles]
{The effects of seeing on S\'ersic profiles. II. The Moffat PSF}
\author[I. Trujillo et al.]
       {I. Trujillo,\thanks{itc@ll.iac.es}$^{1}$ J. A. L. Aguerri,$^{2}$
 J. Cepa$^{1}$ and  C. M. Guti\'errez$^{1}$\\
        $^{1}$ Instituto de Astrof\'{\i}sica de Canarias,  E-38205 La Laguna, Tenerife, Spain\\
	$^{2}$ Astronomisches Institut der Universit\"{a}t Basel, Venusstrasse 7, CH-4102 Binningen, Switzerland}
\date{Accepted 0000 December 00.
      Received 0000 December 00;
      in original form 0000 October 00}

\pagerange{\pageref{firstpage}--\pageref{lastpage}}
\pubyear{2001}

\begin{document}

\maketitle

\label{firstpage}

\begin{abstract}

The effects of seeing on  S\'ersic $r^{1/n}$ profile parameters are extensively
studied using a Moffat function. This analytical approximation to the point
spread function (PSF) is shown to provide the best fit to the  PSF predicted
from  atmospheric turbulence theory when  $\beta\sim4.765$. The  Moffat PSF is
additionally shown to contain the Gaussian PSF as a limiting case
($\beta\rightarrow\infty$). The Moffat function is also shown to be numerically
well behaved when modelling  narrow PSFs  in {\it HST} images. Seeing effects
are computed for elliptically symmetric surface brightness distributions. The
widely used assumption of circular symmetry when  studying the effects of
seeing on intrinsically elliptical sources is shown to produce significant
discrepancies with respect to the true effects of seeing on these sources. A
prescription to correct raw (observed) central intensities, effective radii,
index $n$ and mean effective surface brightness is given. 

\end{abstract}


\section{Introduction}

It is well known that the ability to parameterize  galaxies from ground-based
images is severely compromised by  seeing, which scatters light from the
objects producing a loss of resolution in the images, lower mean surface
brightnesses  than the true values and larger effective radii. The effects of
seeing have been extensively studied in the case of elliptical galaxies with
$r^{1/4}$ profiles (Franx, Illingworth \& Heckman 1989;  Saglia et al. 1993). 
Recently, Trujillo et al. (2001, hereafter T01)  extended previous work by
studying analytically the effects of seeing on elliptically symmetric surface
brightness distributions following the S\'ersic (1968) $r^{1/n}$ law and
assuming a Gaussian point spread function (PSF). S\'ersic's generalization of
the de Vaucouleurs (1948, 1959) $r^{1/4}$  law has been shown to provide a
better representation to the distribution of light in both elliptical galaxies
(including the dwarf ellipticals) and the bulges of Spiral galaxies (Caon,
Cappacioli, \& D'Onofrio 1993; D'Onofrio, Capaccioli, \& Caon 1994; Young \&
Currie 1994; Andredakis, Peletier, \& Balcells 1995). 

The existence of ``wings'' in stellar profiles reveals that the real PSF
deviates from the Gaussian form. In this paper we show, from the size of the
wings present in real images (e.g. Saglia et al. 1993), that such deviations
from Gaussian PSFs can result in different values for the profile parameters in
the range of 10--30\%.  The new generation of ground-based telescopes and the
study of galaxies at high redshifts make these types of studies crucial in
order to obtain reliable (unbiased) information from the structural analysis of
these objects. 

This paper presents a further, more detailed, analysis of the effects of seeing
on S\'ersic profiles when ``wings'' are present in the PSF. For this reason we
have modelled the PSF by a generalization of the Gaussian form: the Moffat
function (Moffat 1969), which  describes well the presence of wings. It should
be noted that these kinds of studies are not only important for  ground-based
observations. In fact, {\it HST} images present their own ``narrow'' PSFs (see
a detailed study in   Bendinelli, Zavatti \& Parmeggiani 1987, and Krist 1993).
The use of these steep PSFs presents numerical problems which can be avoided by
modelling the narrow PSFs with polynomials instead of exponential expressions
like Gaussians. In Section 2 we summarize some general results from the use of
Moffat PSFs. Section 3 describes the effects of seeing on the S\'ersic profile
parameters brought about by the Moffat PSF. A prescription for seeing
corrections is given in Section 4.

\section{General Remarks about Moffat Convolution}

Point spread functions can be determined observationally by studying the
scattering of  stellar light. Numerous papers have been devoted to this problem
(e.g. Moffat 1969; King 1971; Bendinelli et al. 1990). Among the analytical
approximations, the Moffat function (see Eq. [1]) has been widely used to model
the PSF (e.g. Bendinelli, Zavatti \& Parmeggiani 1988a,b; Young et al. 1998);
for instance, the IRAF data reduction package (Tody 1986) adopts the Moffat
function as a standard PSF. In Fig. 1 we plot both the Gaussian function and
Moffat function having a range of $\beta$. Note that as $\beta$ increases the
Moffat function tends to approximate the core of the Gaussian profile. In fact,
a Moffat function contains the Gaussian PSF as a limiting case (see Appendix
A). Moreover, the Moffat PSF has two clear advantages over the Gaussian PSF:

\begin{enumerate}
\item It is numerically well behaved in the treatment of narrow PSFs.
\item It allows the ``wings'' that  usually
appear in stellar profiles to be fitted. 
\end{enumerate}

\begin{figure*}
  \vspace*{200pt}

  \caption{Top panel: The normalized intensity profile of a Gaussian and 
Moffat PSF function having different values of $\beta$ are plotted against the
radius of the PSF in units of FWHM. Bottom panel: The difference between the
normalized Moffat PSF (PSF$_{\beta}$(r)) and  the normalized Gaussian
PSF (PSF$_G$(r)).}

 \label{Fig:NparNprof}
  \end{figure*}

Very accurate convolutions between the PSF and the model profiles of the
galaxies are required in order to obtain reliable results. Current reduction
packages use Fast Fourier Transforms to evaluate the convolutions. This is in
fact inappropriate where there are strong changes in the intensity gradients of
the galaxy profiles. The inner parts of  galaxy profiles are steep, and this
demands a very accurate measurement of the high frequencies in the Fourier
domain. Narrow PSFs (such as those of the {\it HST\/}) magnify this problem
because they also present a steeper profile. Working in the {\it real} domain
does not exempt us from trouble either; in fact, one can encounter  several
numerical problems when performing accurate convolutions when using a Gaussian
to model narrow PSFs. Current computers can manage numbers of the order of
$\sim e^{200}$. These kinds of numbers can be easily obtained when working with
Gaussians which have $\sigma<1$ in units of pixels (see the exponential
expressions at play when performing a Gaussian convolution in Eq. [4] in T01). 
The use of Moffat functions avoids this problem due to the use of polynomials 
instead of exponential expressions (see Eq. [3]). In this sense, Moffat
functions are numerically better behaved than Gaussians when dealing with
narrow PSFs.

\subsection{Mathematical analysis}

We will use a circular Moffat function to model the point spread function:

\begin{equation}
{\rm PSF}(r)=\frac{\beta-1}{ \pi \alpha^2}
\left[ 1+\left(\frac{r}{\alpha}\right)^2\right]^{-\beta},
\end{equation}
with the full width at half maximum, FWHM=$2\alpha\sqrt{2^{1/\beta}-1}$, 
where PSF(FWHM/2) = (1/2)PSF(0) and the
total flux is normalized to 1.
Consider a case where, in the absence of seeing, the surfaces brightness 
distribution, $I({\bf {r}})$, of a galaxy is elliptically symmetric. This
means  that the isophotes of the object all have the same constant ellipticity 
$\epsilon$ ($\epsilon=1-b/a$, where $a$ and $b$ are respectively the semi-major  and
semi-minor axes of the isophotes).

As shown in T01 elliptical coordinates ($\xi,\theta)$ are the most appropriate
for this type of  problem. In this coordinate system,  the surface brightness
distribution, $I({\bf {r}})$, of  an elliptical source depends only on $\xi$:
$I({\bf {r}})=I(\xi)$.  The convolution equation that represents the effect of
seeing on the  surface brightness distribution is given by:

\begin{equation}
I_c(\xi,\theta) = (1-\epsilon) \int_0^\infty \xi^{'} I(\xi^{'})d\xi^{'} \int_0^{2 \pi} 
d\theta^{'} {\rm PSF}(\xi^{'}, \theta^{'}, \xi, \theta) ,
\end{equation}
where PSF$(\xi^{'}, \theta^{'}, \xi, \theta)$ is the Moffat PSF given by:

\begin{eqnarray}
\lefteqn{{\rm PSF}(\xi^{'}, \theta^{'}, \xi, \theta)=\frac{\beta-1}{\pi
\alpha^2}\bigg[ 1+ {}}
\nonumber \\
& & {}
\frac{\xi^2+\xi^{'2}-2\xi\xi{'}\cos(\theta-\theta^{'})+(\epsilon^{2}-2\epsilon)(\xi^{'}
\sin\theta^{'}-\xi \sin\theta)^2}{\alpha^2}\bigg]^{-\beta}.
\end{eqnarray}

The subscript ``c'' shall be used from here on to refer seeing--convolved
quantities. Along the major axis of the object, $\theta$=0, the angular integral
can be  solved analytically ($\epsilon>$0):

\begin{eqnarray}
\lefteqn{
\int_0^{2 \pi}  d\theta^{'} {\rm PSF}(\xi^{'}, \theta^{'}, \xi,0)=2
\frac{\beta-1}{\pi \alpha^2} a_{\epsilon}^{-\beta}
\sum_{k=0}^{\infty}C_{2k}^\beta(w) {}}
\nonumber \\
& & {}
\left[\frac{1}{a_{\epsilon}}\left(\frac{\xi^{'}}{{\alpha}}
\right)^2 (2\epsilon-\epsilon^{2})\right]^{k}B\left(\frac{1}{2},
\frac{2k+1}{2}\right),
\end{eqnarray}
where
\begin{equation}
 a_{\epsilon}\equiv1+\frac{\xi^{'2}(1-\epsilon)^2+\xi^2}{\alpha^2}
 \mbox{ and }w\equiv\frac{1}{(2\epsilon-\epsilon^{2})^{1/2}}\frac{\xi}{\alpha}
 \frac{1}{a_{\epsilon}^{1/2}}
\end{equation}
and $C_n^\lambda(t)$ and $B(z,w)$ are the Gegenbauer polynomials (Gradshteyn \&
Ryzhik 1980, p. 1029) and beta functions  (Abramowitz \& Stegun 1964, p. 258)
respectively.  A simpler expression is obtained for the convolution in the
circularly symmetric case ($\epsilon=0$):

\begin{equation}
I_{\rm c}(r)=2\frac{\beta-1}{\alpha^2} \int_0^\infty dr^{'}r^{'}
I(r^{'})a_0^{-\beta}P_{\beta-1}\left[ a_0^{-1}
\left(1+\frac{r^2+r^{'2}}{\alpha^2}\right)
\right],
\end{equation}
where 
\begin{equation}
a_0\equiv\left[\left(1+\frac{r^{'2}-r^2}{\alpha^2}\right)^2+\left(\frac{2r}{\alpha}\right)^2\right]^{1/2}
\end{equation}
and $P_n(x)$ is the Legendre function of first class (Abramowitz
 \& Stegun 1964, p. 332).
\subsection{The effect of seeing on the central intensity}

For any  intensity distribution with elliptical symmetry, $I(\xi)$,
the seeing convolved central intensity, $I_{\rm c}(\xi=0)$, is such that

\begin{equation}
I_{\rm c}(0)=2\frac{\beta-1}{\alpha^2}(1-\epsilon)\int_0^\infty
d\xi^{'}\xi^{'}I(\xi^{'})\frac{1}{(b^2-c^2)^{\beta/2}}P_{\beta-1}
\left(\frac{b}{\sqrt{b^2-c^2}}\right),
\end{equation}
where
\begin{equation}
b\equiv1+\frac{\xi^{'2}}{\alpha^2}\left[ 1+\frac{1}{2}(\epsilon^2-2\epsilon)\right] \mbox{
and }  c\equiv-\frac{\xi^{'2}}{2\alpha^{2}}(\epsilon^2-2\epsilon).
\end{equation}

Note that the effect on the central intensity of the seeing convolved
distribution is a function of the intrinsic ellipticity of the object.
Basically, as $\epsilon$ increases the spread of photons from the inner parts
of the profile due to the seeing is more efficient and consequently the central
intensity decreases.

\subsection{The effect of seeing on the
ellipticity of the isophotes}

In the absence of seeing, by construction, all isophotes of the  profile have
the same ellipticity, whereas the presence of seeing tends to make them
circular. Using the isophote condition, $I_{\rm c}(\xi,0)=I_{\rm c}(\xi,\pi/2)$ it is possible to
derive  an implicit equation that gives the  variation of the ellipticity with
the radial distance:

\begin{eqnarray}
\lefteqn{
 \int_0^\infty \xi^{'}
I(\xi^{'})d\xi^{'}\sum_{k=0}^{\infty}\left[\frac{C_{2k}^\beta(w)}{a_\epsilon^{\beta+k}}-(-1)^k
\frac{C_{2k}^\beta(w^*)}{(a_{\epsilon}^*)^{\beta+k}}\right] {}}
\nonumber \\
& & {}
\left[(2\epsilon-\epsilon^{2})\left(\frac{\xi^{'}}{{\alpha}} \right)^2
\right]^{k}B\left(\frac{1}{2},\frac{2k+1}{2}\right)=0.
\end{eqnarray}

For this problem, it is useful to introduce  $a_\epsilon, w,
 a_{\epsilon}^*$ and $w^*$  as functions of the Cartesian
 coordinates $x,y$:
\begin{equation}
 a_{\epsilon}\equiv1+\frac{\xi^{'2}(1-\epsilon)^2+x^2}{\alpha^2}
 \mbox{ and }w\equiv\frac{1}{(2\epsilon-\epsilon^{2})^{1/2}}\frac{x}{\alpha}
 \frac{1}{a_{\epsilon}^{1/2}}
\end{equation}
and
\begin{equation}
 a_{\epsilon}^*\equiv1+\frac{\xi^{'2}+y^2}{\alpha^2}
\mbox{ and }w^*\equiv-i\frac{1}{(2\epsilon-\epsilon^{2})^{1/2}}
\frac{y}{\alpha} \frac{1}{(a_{\epsilon}^*)^{1/2}}.
\end{equation}
From this implicit equation we can obtain $y/x$ and therefore the ellipticity
of the isophotes affected by seeing using $\epsilon(x)=1-y/x$.

\subsection{The ability of the Moffat function to match the atmospheric
turbulence  prediction of the PSF}

The theory of atmospheric turbulence predicts the PSF to be the Fourier
transform of exp[$-(kb)^{5/3}$] (Fried 1966; Woolf 1982), where FWHM =
2.9207006$b$ and $b$ is a scaling parameter. In the {\it real} domain this PSF
is written as:

\begin{equation} {\rm PSF}_{T}(r)=\frac{1}{2\pi} \int_0^\infty k
J_0(kr)e^{-(k\frac{\rm FWHM}{2.9207})^{5/3}}dk, \end{equation} 
where $J_0$ is the standard Bessel function (Abramowitz \& Stegun 1964, p.
358). For a given FWHM, we have evaluated the value of $\beta$ that minimizes
the difference between the prediction of the atmospheric turbulence theory and
the Moffat function by minimizing the $\chi^2$ of the fit between both PSFs. 
An optimum value of $\beta\sim 4.765$ was found. In Fig. 2 we have shown the
difference between the PSF prediction from  turbulence theory and a Moffat
function for a value of $\beta=4.765$. It can be seen that the agreement is
quite good. A Moffat function could therefore  be used to reliably model the
turbulence prediction, although, the PSFs usually measured in  real images
have bigger ``wings'', or equivalently smaller values of $\beta$, than those
expected from the turbulence theory (e.g. Saglia et al. 1993). This is because
the real seeing not only depends  on  atmospheric conditions but is also caused by
imperfections in telescope optics.   

The presence of these bigger ``wings'' in real images makes Moffat functions a
better choice to model the PSF than the turbulence theory prediction. As an
example of this, current packages of data reduction in IRAF suggest a default
value of $\beta=2.5$. In order to span the  range of the different ``wing''
sizes present in real images, in the next section we have modelled the Moffat
PSFs  using three different values of $\beta$: $\beta$=5 (to simulate the
turbulence prediction), 2.5 (the default value of the IRAF package), and 1.5
(to model a large ``wing'' in the PSF). 

\begin{figure*}
  \vspace*{200pt}

  \caption{Top panel: The best fit to the PSF predicted by the theory of 
atmospheric turbulence (diamonds) using a Moffat function with a $\beta$ value of
4.765 (solid line). Bottom panel: The difference between these two PSFs as a function of
the radius of the PSF in unit of FWHM.}

\label{} \end{figure*}

\section[]{The effects of seeing on the S\'ersic profile parameters} 

The equations that we have shown in the previous section are general results
for  Moffat seeing. For practical purposes with applications to real galaxies,
we are going to focus on the S\'ersic profile. In the particular case of
$r^{1/n}$ profiles, the surface brightness distribution is given (in elliptical
coordinates) by: 

\begin{equation}
I(\xi)=I(0)10^{-b_n\left[(\xi/r_{\rm e}\right)^{\frac{1}{n}}]},
\end{equation}
where $I(0)$ is the central intensity and $r_{\rm e}$ 
the effective radius of the
profile. The constant $b_n$ is chosen such that half the total luminosity predicted by
the law comes from  $\xi<r_{\rm e}$. $b_n$ can be well approximated by the relation 
$b_n=0.868n-0.142$ (Caon et al. 1993).

\subsection[]{The central intensity}

To study the effect of seeing on the central intensity we make use of Eqs. (8) and
(14). Figure 3 shows this effect for different values of the ellipticity in the 
intrinsic light profile,
assuming $\beta$ = 1.5, 2.5, 5 and for the Gaussian case (i.e. $\beta \to \infty$) as well.

na As $\beta$
increases (i.e. as the size of the PSF ``wings'' decrease) we asymptotically
recover the effect on the central intensity  produced by a Gaussian PSF.

From this figure it follows  that the effect of  seeing on the central
intensity increases as the size of the ``wings'' increase. This is easily
understood as broader ``wings'' increase the probability that a photon will
hit the imaging device at a point  further offset from where it would
have hit in the absence of a seeing. For a typical value of $\beta$ (e.g. $\beta$ = 2.5),
the difference from  a Gaussian PSF is $\sim$ 10\%.

 \begin{figure*} 
\vspace*{200pt}

  \caption{The effects of seeing on the observed central intensity, $I_{\rm
c}(0)$, for different values of $n$. The Gaussian case (solid line) and three
different values of $\beta$ are shown:   $\beta=5$ (dot-dashed line),
$\beta=2.5$ (dashed line) and $\beta=1.5$ (dot-dot-dot-dashed line). Three
different ellipticities are also shown: $\epsilon=0$ (left column),
$\epsilon=0.25$ (middle column) and $\epsilon=0.5$ (right column).}  

\label{}  
\end{figure*}

For a given seeing--disc (n.b.\ the relative size of the FWHM to effective
radius {\it increases} along the x--axis of Fig. 3, c.f.\ Fig. 4 and Fig. 5),  
the central intensity of  profiles with larger values of $n$ is more affected
than for low $n$ as is expected because of the higher central light
concentration of these profiles. As noted in Section 2.2, the  effect of seeing
on the central intensity of the object is also dependent on the intrinsic
ellipticity of the object: the central intensity of galaxies with larger
ellipticities are more affected by seeing. 

\subsection[]{The effective radius}

The seeing effect on effective radius can be obtained by solving for $r_e^c$
from  the conservation of luminosity by the convolution $L^{\rm c}(r_{\rm
e}^{\rm c})$=$L(r_{\rm e})$, where $L(r_e)$ is the luminosity of the source
inside $r_{\rm e}$  and  $L^{\rm c}(r_{\rm e}^{\rm c})$ is  the luminosity
obtained from the object affected by seeing, measured inside  its effective
radius.

Figure 4 shows this effect for different values of ellipticity, with $\beta$ =
1.5, 2.5, 5 and for the Gaussian PSF. Here, the presence of significant
``wings'' in the PSF produces an effective radius larger than what is expected
from  Gaussian seeing. As $n$ increases the {\it convolved} effective radius
also increases. The ellipticity effect is also shown. Greater ellipticities
result in greater effective radii, and these differences are more important for
greater values of $n$. This result is as expected due to the diminution of the
central intensity with larger  ellipticity. For the values of $\beta$
typically present in real images (2.5 $<\beta<$ 4; see Saglia et al. 1993) we
obtain deviations from Gaussian seeing in the range 15--30\% (bigger deviations
are obtained for smaller values of the ratio $r_{\rm e}^{\rm c}/{\rm FWHM}$).

\begin{figure*} 

\vspace*{200pt}

  \caption{The effects of seeing on the observed effective radius, $r_e^c$ for
different values of $n$. The Gaussian case (solid line) and three different
values of $\beta$ are shown: $\beta=5$ (dashed line), $\beta=2.5$ (dotted line) and
$\beta=1.5$ (dot-dashed line). Three different ellipticities for the sources
are also shown, $\epsilon=0$ (left column), $\epsilon=0.25$ (middle
column) and $\epsilon=0.5$ (right column).}  

  \label{}
\end{figure*}

It should be noted that our measurement of the effective radius has  been
obtained over the semi-major axis. Some authors use as radial  distance the
magnitude $r^*=\sqrt{ab}$; in this case, the effective  radius of the object
affected by seeing is given by  $r_{\rm e}^{\rm c*}=
r_{\rm e}^{\rm c}\sqrt{1-\epsilon (r_{\rm e}^{\rm c})}$,
where $\epsilon(r_{\rm e}^{\rm c})$  can be obtained using Equation (10).

\subsection{The S\'ersic index {\boldmath $n$} }

To quantify the effect of seeing on the shape parameter $n$ we use the
parameter $\eta(\xi)$ (T01). $\eta(\xi)$ is equivalent, locally, to the
parameter $n$ of the S\'ersic profile. This parameter can be understood as a
measure of the slope of the profile. In Fig. 5 we show the effects of seeing on
this parameter (evaluated at  $r_{\rm e}^{\rm c}$) for different sizes of the
PSF  ``wings'' and different intrinsic ellipticities. It is easy to see how the
real value of $n$ is recovered asymptotically when the ratio   $r_{\rm e}^{\rm
c}/{\rm FWHM}$ increases. The effect of seeing on $\eta(r_{\rm e}^{\rm c})$ is
bigger for smaller values of $\beta$. This means that bigger PSF ``wings''
produce a stronger effect on the slope of the profile. The increase in the
intrinsic ellipticity of the profile has a similar effect, but the influence is
not  as important as it was for the previous parameters. Note that seeing
effects always produce a surface brightness profile with a smaller value of $n$
than the actual one. It is expected that any procedure to recover the
structural parameters of the profile that does not take into account the effect
of seeing will obtain lower values of $n$ than the actual ones. Lower values of
$n$  will also be expected  if the intrinsic ellipticities of the objects are
not taken into account during the recovery process. This is crucial in the
study of high--z galaxies. As  was shown for the central intensity and
effective radius, the presence of ``wings'' in the PSF causes the values of the
profile parameters to deviate from the  prediction made correcting for Gaussian
seeing. In the case of the $\eta(r_{\rm e}^{\rm c})$ parameter, these
deviations are   in the range of 10--20\%.

\begin{figure*} 

%

\vspace*{200pt}

  \caption{The effects of seeing on the index  $\eta(r_{\rm e}^{\rm c})$ for different
values of $n$. The Gaussian case (solid line) and three different
values of $\beta$ are shown: $\beta=5$ (dashed line), $\beta=2.5$ (dotted line) and
$\beta=1.5$ (dot-dashed line). Three different ellipticities for the sources
are also shown, $\epsilon=0$ (left column), $\epsilon=0.25$ (middle column) and
$\epsilon=0.5$ (right column).}  

  \label{}
\end{figure*}

\subsection{The mean effective surface brightness}

From the previous results it is now easy to study the effect of  seeing on
the mean effective surface brightness,  defined as: 

\begin{equation}
\langle\mu\rangle _{\rm e}\equiv-2.5 \log \frac{L(r_{\rm e})}{\pi r_{\rm e}^2}.
\end{equation}
By using Eq. (15) and the conservation of the flux, it  immediately follows
that

\begin{equation}
\Delta \langle\mu\rangle _{\rm e}\equiv\langle\mu\rangle _{\rm e}^{\rm c}
-\langle\mu\rangle _{\rm e} =5 \log \frac{r_{\rm e}^{\rm c}}{r_{\rm e}}.
\end{equation}
\begin{figure*}
  \vspace*{200pt}

  \caption{The differences $\Delta\langle\mu\rangle_{\rm e}$ between the measured mean
effective surface brightness, $\langle\mu\rangle_{\rm e}^{\rm c}$ and the seeing-free quantity
$\langle\mu\rangle_{\rm e}$ for different values of $n$. 
The Gaussian case (solid line) and three different
values of $\beta$ are shown: $\beta=5$ (dashed line), $\beta=2.5$
(dotted line) and $\beta=1.5$ (dot-dashed line). Three different ellipticities
for the sources are also shown, $\epsilon=0$ (left column), $\epsilon=0.25$
(middle column) and $\epsilon=0.5$ (right column).}  

  \label{}
\end{figure*}

In Fig. 6 we show  how the mean effective surface brightness changes as a
function of $r_{\rm e}^{\rm c}/$FWHM for different values of $n$ and intrinsic
ellipticities. This figure clearly shows that galaxies affected by seeing have
apparent mean surfaces brightnesses lower than their true values. Lower values
of $\beta$ produce greater effects on this quantity. Also, as the  intrinsic
ellipticity of the object increases the effects of the seeing on the mean
effective surface brightness also increase. 

\section[]{A prescription for seeing corrections}

It is possible to obtain the parameters of the S\'ersic profiles (seeing-free
quantities) from the convolved quantities. We present an easy
prescription\footnote{A similar prescription for  Gaussian seeing is presented
in T01.} based on the use of the plots of Figures 3, 4 and  5. The steps which
observer must follows are:

\begin{itemize} 

\item Determine the FWHM and the value of $\beta$ from the stellar profile by
fitting a Moffat function. Current astronomical data reduction packages, such
as IRAF, allow this fitting.

\item Measure $r_{\rm e}^{\rm c}$ along the semi-major axis directly from the raw images. This
can be done by solving the implicit equation 
$L^{\rm c}(r_{\rm e}^{\rm c})=(1/2)L^{\rm c}(\infty)$.

\item Determine $\eta(r_{\rm e}^{\rm c})$ numerically using the expression: 
\begin{equation}
\eta(r_{\rm e}^{\rm c})=\frac{1}{r_{\rm e}^{\rm c}}\frac{I_{\rm c}(r_{\rm e}^{\rm c})}
{\frac{dI_{\rm c}(\xi)}{d\xi}\arrowvert_{r_{\rm e}^{\rm c}}} \ln \frac{I_{\rm c}(r_{\rm e}^{\rm c})}{I_{\rm c}(0)}.
\end{equation}

\item Evaluate the value of $n$ and $\epsilon$ from the use of Figure 5.

\item Obtain the value of $r_{\rm e}$ using Figure 4. 

\item Obtain the value of $I(0)$ using Figure 3.
   
\end{itemize}

\section{Conclusions}

As redshift increases, the apparent size of  galaxies becomes progressively
smaller and the affect of seeing progressively stronger. As we have seen,
precise corrections for seeing are demanded in order to obtain reliable and
comparable information  about structural parameters from objects at the same or
different redshifts. We have chosen the Moffat function to model the PSFs of
real images. This choice has been made by following two criteria: the ability
of this function to model the ``wings'' of the PSFs present in  real images
obtained from ground-based telescopes, and it is well behaved numerically
because of its polynomial structure.

We have  studied the general properties of the Moffat function when 
modelling  PSFs. These properties can be summarized as follows: it is a very
good option to model the narrow PSFs present, for example, in  {\it HST} images
because it is numerically well behaved; the Gaussian PSF is a limiting case of
the Moffat PSF ($\beta\rightarrow\infty$); and the prediction for the PSF due
to the theory of atmospheric turbulence can be numerically well approximated by
a Moffat function with $\beta\sim4.765$. 

For practical purposes we have analysed the effects of seeing caused by this
PSF on the S\'ersic model. The effects on the central intensity, effective
radius,  $n$ index and mean effective surface brightness are extensively shown
in Figures 3, 4, 5 and 6. We  have also given an easy prescription for seeing
correction that can be useful for  observers in order to obtain the seeing-free
quantities.

Our main results have been to show the importance of taking into account the
intrinsic ellipticities of the objects and the presence of ``wings'' in the
PSFs for the recovery of accurate structural parameter. It is not sufficient to
consider the PSF as Gaussian and assume circular symmetry to model the effects
of seeing on the surface brightness distribution when the ratio of the effective
radius to the FWHM is small ($\leq 2.5$).

\section*{Acknowledgments}

We wish to thank Alister W. Graham who kindly proofread versions of this
manuscript.

\appendix

\section[]{The Gaussian as a limiting case of the Moffat function}

The Gaussian function can be obtained from the Moffat function (MF) in the
limiting case where $\beta\rightarrow\infty$. One can rewrite the MF as a
function of FWHM  ($F$) and $\beta$:

\begin{equation}
{\rm PSF}(r)=4\left( 2^{1/\beta}-1\right)\frac{\beta-1}{ \pi F^2}
\left[1+4(2^{1/\beta}-1)\left(\frac{r}{F}\right)^2\right]^{-\beta}.
\end{equation}
As $\beta\rightarrow\infty$, we can substitute $2^{1/\beta}-1$ with
${(\ln2)}/{\beta}$, so:
\begin{equation}
\lim_{\beta\rightarrow\infty}{\rm
PSF}(r)=\lim_{\beta\rightarrow\infty}\frac{\beta-1}{\beta}\frac{4\ln2}{ \pi
F^2} \left[1+\frac{4\ln2}{\beta}\left(\frac{r}{F}\right)^2\right]^{-\beta}.
\end{equation}
Using $\lim_{m\rightarrow\infty}(1+{z}/{m})^m=e^{z}$ we have
\begin{equation}
\lim_{\beta\rightarrow\infty}{\rm
PSF}(r)=\frac{4\ln2}{ \pi
F^2}e^{-\frac{4\ln2}{F^2}r^2}.
\end{equation}
Finally, writing $F^2=8\sigma^2\ln2$ we obtain:
\begin{equation}
\lim_{\beta\rightarrow\infty}{\rm
PSF}(r)=\frac{1}{2\pi\sigma^2}e^{-\frac{1}{2}\left(\frac{r}{\sigma}\right)^2}.
\end{equation}
For practical purposes, a value of $\beta=100$ is
completely satisfactory for modelling a Gaussian by using a Moffat function.

\bsp

\label{lastpage}

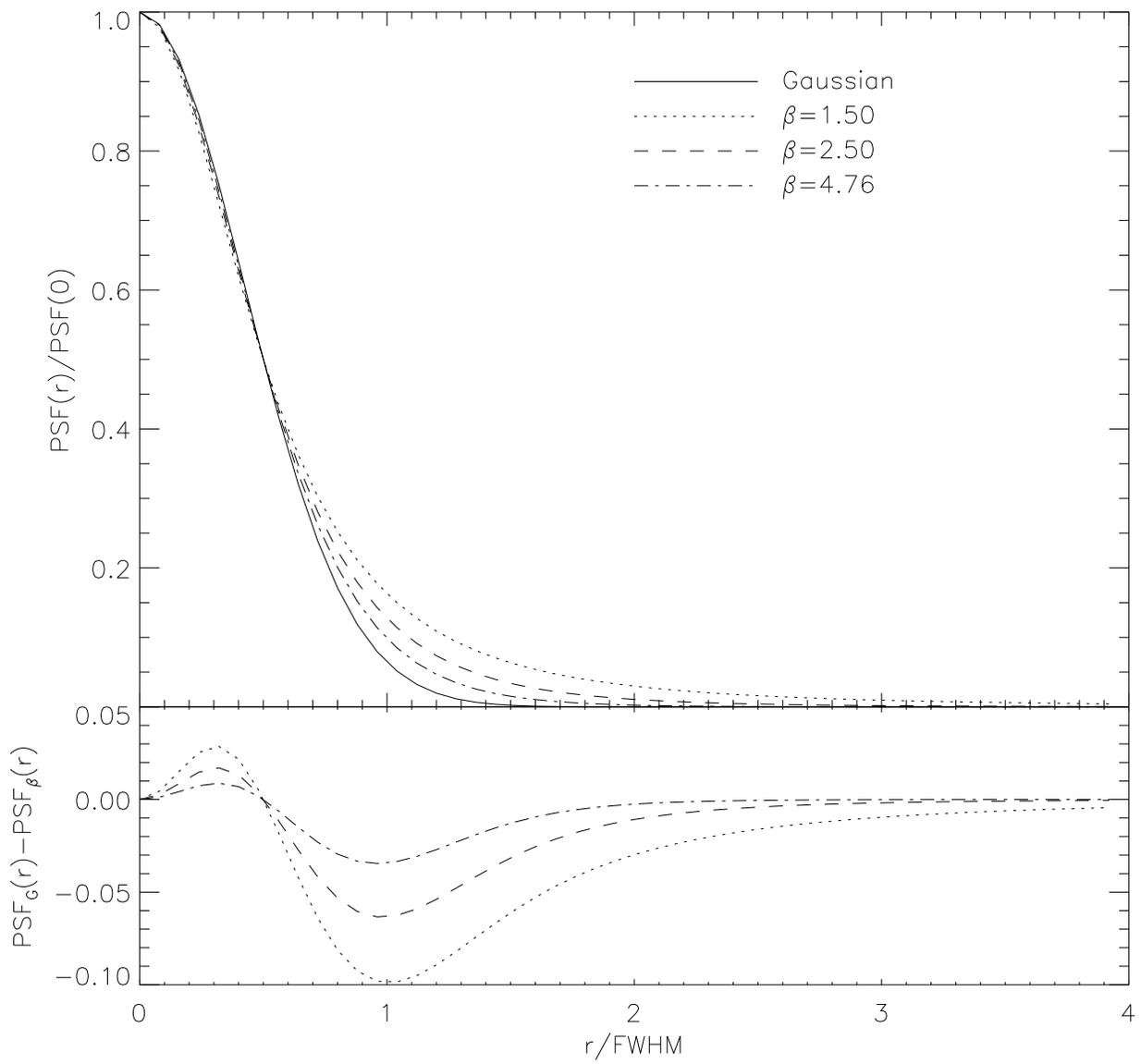
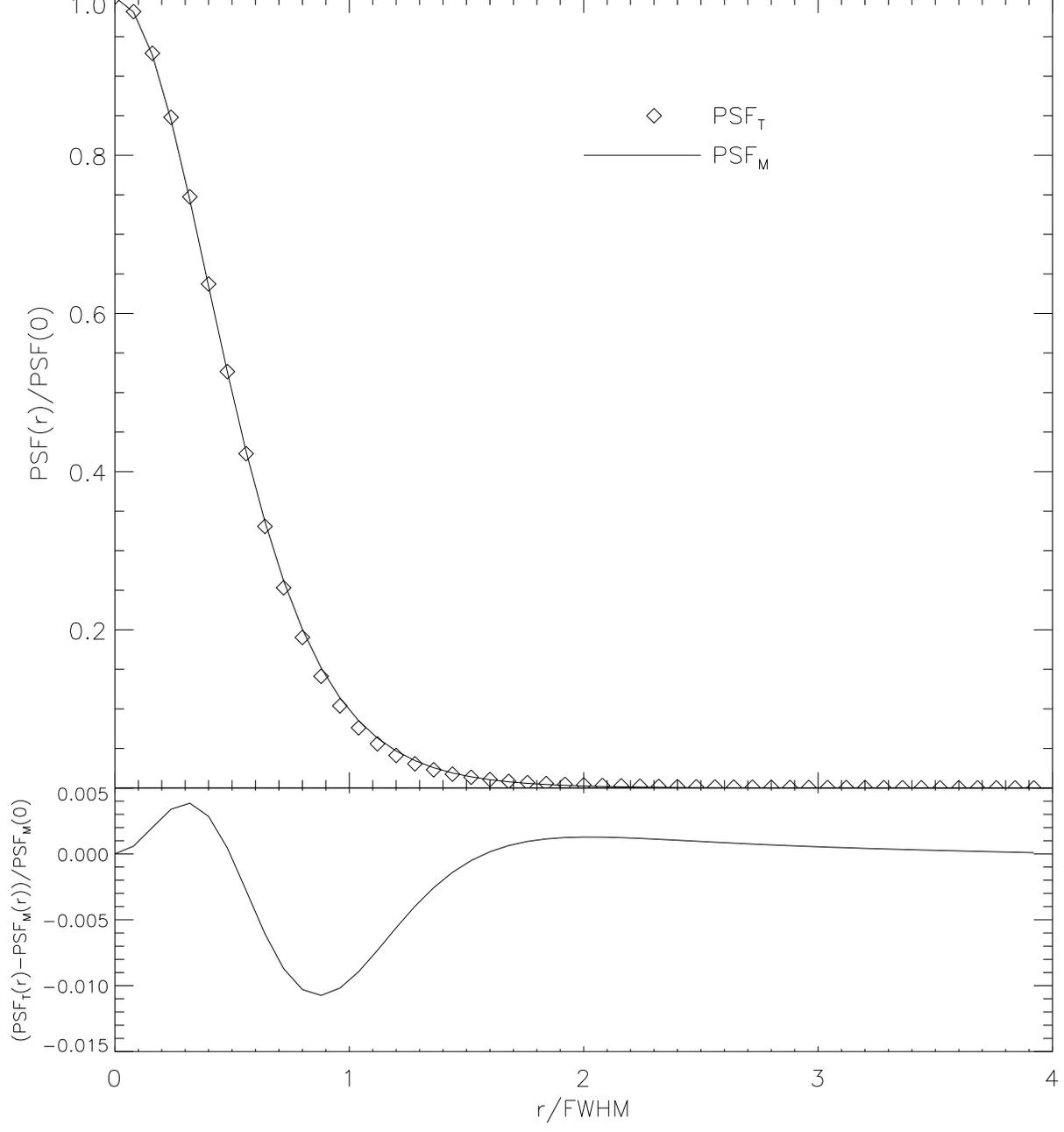
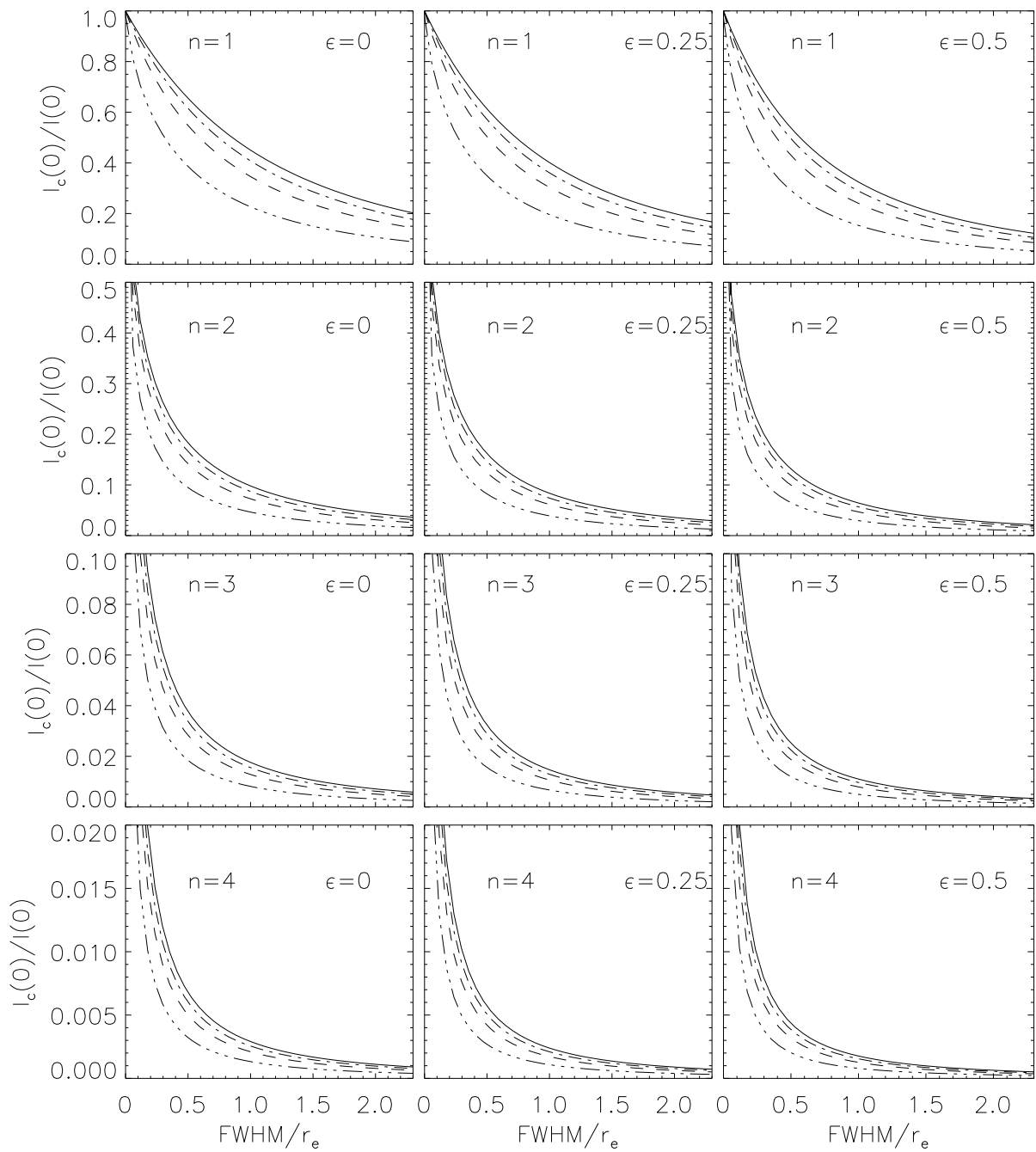
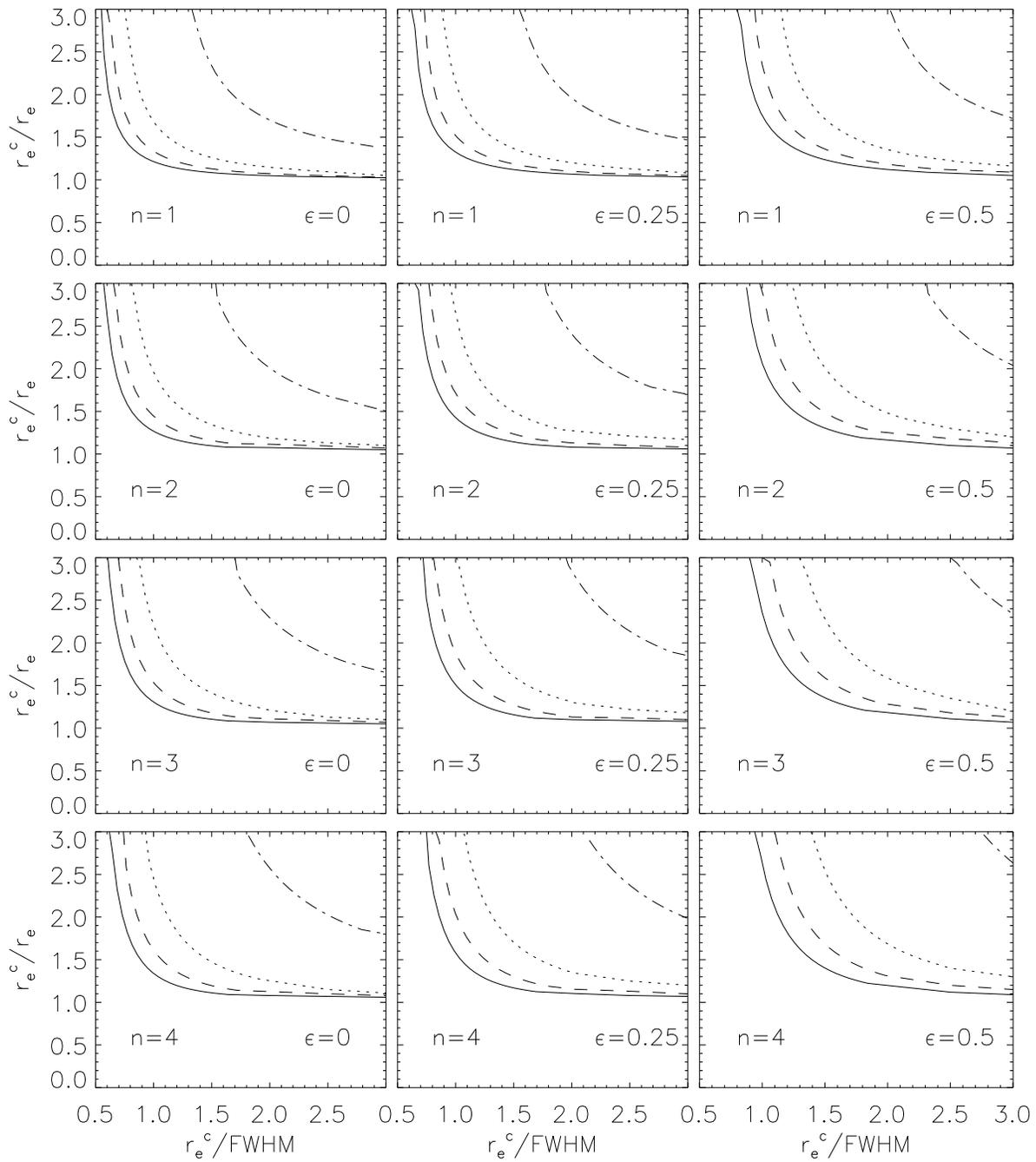
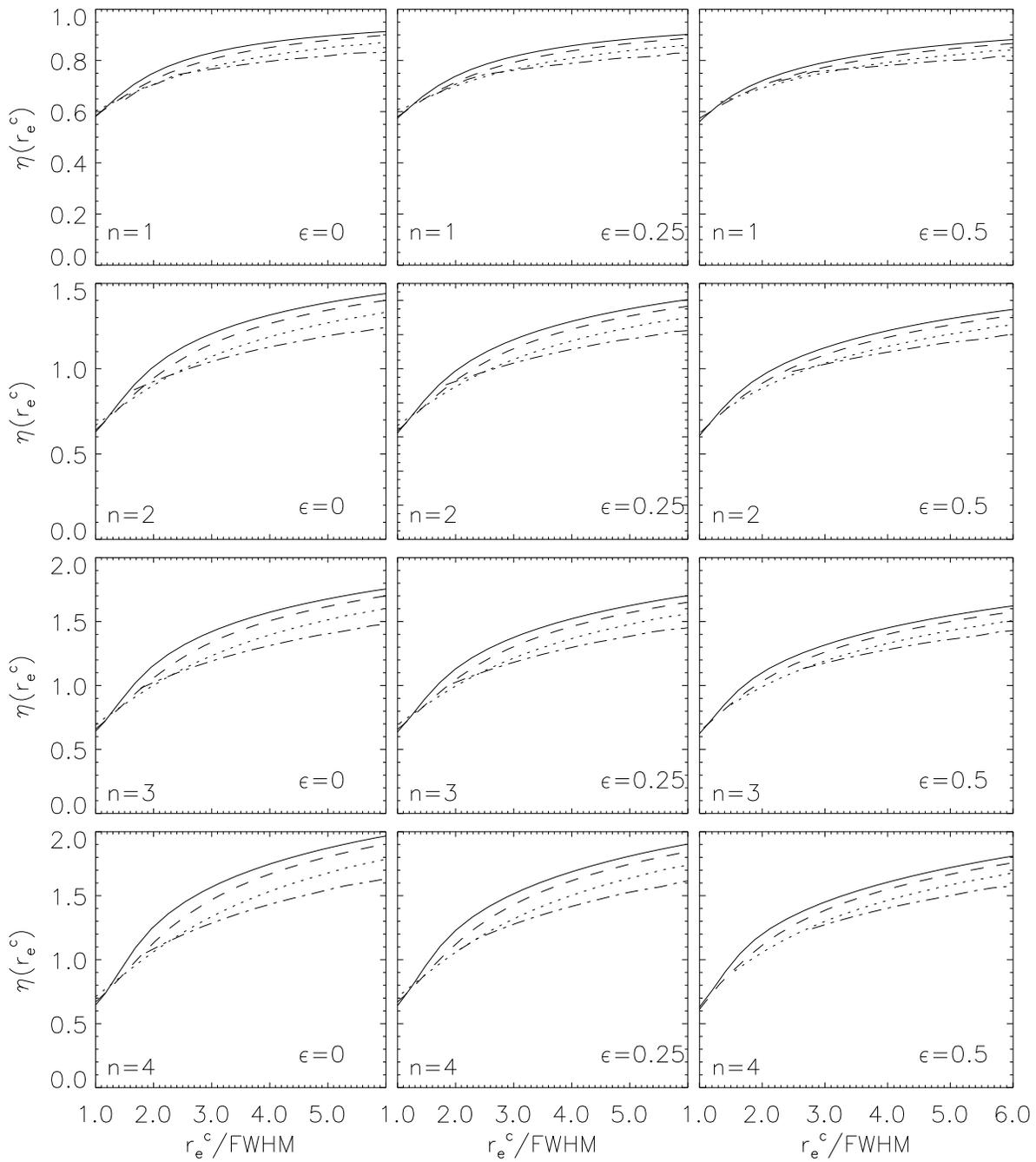
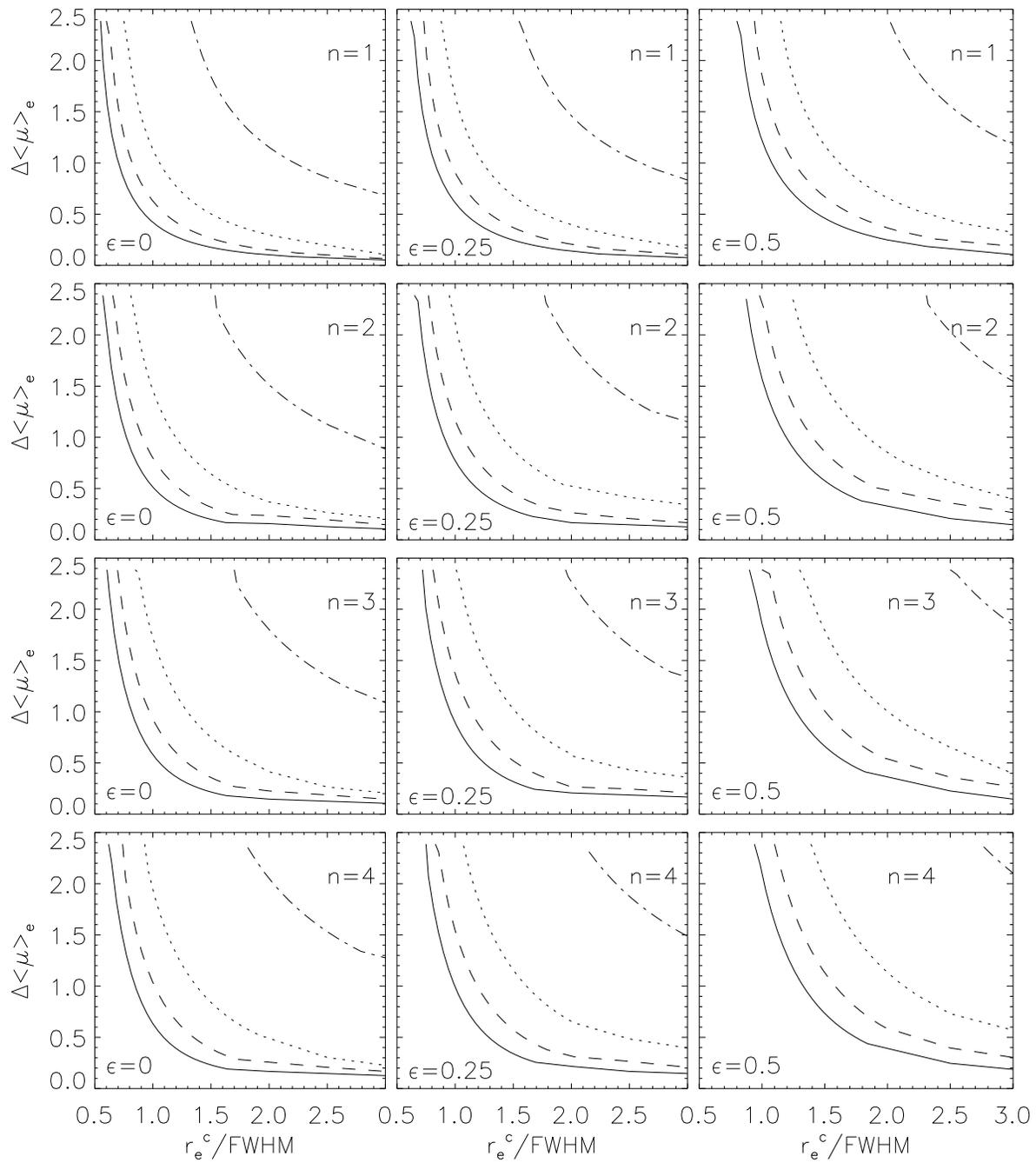
\end{document}